\documentclass []  {aa}

\usepackage{graphicx}

\begin{document}

\title{ Photometric and spectroscopic study of  low mass embedded star
clusters in reflection nebulae\thanks{Based on observations made at
Complejo  Astron\'omico El Leoncito, which is operated under agreement
between the  Consejo Nacional de Investigaciones Cient\'{\i}ficas y
T\'ecnicas de la  Rep\'ublica Argentina and the National Universities
of La Plata, C\'ordoba  and San Juan, Argentina.} }

\author{J. B. Soares \inst{1}, E. Bica \inst{1}, A. V. Ahumada
\inst{2}, J.  J. Clari\'a  \inst{2}}

\offprints{ J. B. Soares - jules@if.ufrgs.br}

\institute{Universidade Federal do Rio Grande do Sul, IF, CP\,15051,
Porto  Alegre 91501--970, RS, Brazil \and Observatorio Astron\'omico
de C\'ordoba,  Laprida 854, 5000, C\'ordoba, Argentina}

\date{}

\abstract {  An analysis of the candidate embedded stellar systems
in the reflection  nebulae vdBH-RN\,26, vdBH-RN\,38, vdBH-RN\,53a,
GGD\,20, ESO\,95-RN\,18 and  NGC\,6595 is presented.  Optical
spectroscopic data from CASLEO (Argentina) in conjunction with near
infrared photometry from the 2MASS Point Source Catalogue were
employed. The  analysis is based on source surface density,
colour-colour and  colour-magnitude diagrams together with theoretical
pre-main sequence  isochrones. We take into account the field
population affecting the analysis by carrying out a statistical
subtraction. The fundamental parameters for the stellar systems were
derived.  The resulting ages are in the range 1-4 Myr and the objects
are dominated by  pre-main sequence stars. The observed masses locked
in the clusters are  less than  25$M_{\odot}$. The studied systems
have no stars of spectral types  earlier than B, indicating that star
clusters do not necessarily evolve  through an HII region phase. The
relatively small locked mass combined with the fact that they are not
numerous in catalogues suggests that these low mass clusters are not
important donors of stars to the field populations. 
 } 
 \titlerunning { Photometric and spectroscopic study of
embedded clusters in nebulae} \authorrunning {J. B. Soares et al. }
\maketitle

\section{Introduction}

The sensitivity increase of infrared detectors has provided insight
into the  understanding of early evolutionary stages of stars and
stellar clusters.  Molecular clouds have revealed a surprisingly large
number of embedded star  clusters, indicating that a significant
fraction, if not the vast majority  of all stars, may form in such
systems (Lada \& Lada 2003).

Embedded clusters emerge from molecular clouds owing to the disruptive
action of newly born stars, including the less massive ones (Matzner
\&  McKee 2000). This process limits the efficiency of star formation
from gas,  decreasing gravitational energy and resulting in a large
mortality rate for  embedded clusters (EC). Approximately 95\% of the
embedded clusters emerge from  molecular clouds as unbound objects and
the systems that evolve to stable  open clusters have masses $M_{EC} >
500M_{\odot }$ (Kroupa \& Boily 2002).

Recent surveys of infrared star clusters and stellar groups are based
on  directions of optical and radio HII regions (e.g. Dutra et
al. 2003) or  Herbig AeBe stars (e.g. Testi et al. 1998). Dutra et
al. (2003) and Bica et  al. (2003b), who investigated more than 3000
optical and radio nebulae with  2MASS  (Skrutskie et al. 1997), found
346 new embedded star clusters, stellar groups and candidates.  All
together, these surveys provide a large sample for detailed studies.

 The main goals of the present work are to further study young
 star clusters or groups related to reflection nebulae, and extract
 their physical parameters by means of a series of tools, such as
 density profiles,  colour-colour and colour-magnitude diagrams.  We
 intend to provide insight to the understanding of their evolution as
 compared to those of clusters with massive stars producing HII
 regions. A closeby similar object is the sparse young  cluster in the
 reflection nebula NGC\,5367 (Williams et al. 1977). It is  important
 to explore other objects of this type and they are not necessarily
 closeby. Testi et al. (1998)  found  sources around Herbig AeBe
 stars, part  of them related to reflection nebulae.  Soares \& Bica
 (2002, 2003) studied four low mass star clusters in the CMaR1
 molecular cloud, three of them related to optical reflection nebulae.

  For a broader view we analyse in the present study some candidates to
clusters in reflection nebulae. Most of the  clusters and stellar
groups in  Bica et al. (2003b) and Dutra et al. (2003)  are massive
ones  related to optical and radio HII regions and their molecular
clouds.  The subsample of star clusters and groups in those studies
related to optical reflection nebulae is small, which makes them
interesting to explore.  

The reflection nebulae  vdBH-RN\,26, vdBH-RN\,38 and vdBH-RN\,53a
were  catalogued by van den Bergh \& Herbst (1975), ESO\,95-RN\,18 by
Lauberts (1982) and GGD\,20  by Gyulbudaghian et
al. (1978).  The latter  object
is equivalent to BBW\,311 (Brand et al. 1986).   Van den Bergh \&
Herbst (1975) provided a distance from the Sun  d\( _{\odot } \)~$ =
1.7$ kpc for  vdBH-RN\,26.  Herbst (1975) reported  for vdBH-RN\,38, {\bf
 the core of the nebula ESO\,128EN\,4,}
d\( _{\odot } \)~$ = 2.5$ kpc  from UBV photometry of the central
star, but no spectroscopy was available.  Brand \& Blitz (1993)
obtained for the nebula BBW\,311 d\( _{\odot } \)~$ =  3.1$
kpc. vdBH-RN\,53a is equivalent to ESO\,94?6 and BBW\,371A, with
$\approx0.6\arcmin$. Two other small nebulae are in the area. Herbst
(1975)  obtained for vdBH-RN\,53a {\bf d\( _{\odot } \)~$ = 4.2$}
kpc. Brand \& Blitz  (1993) found for the general direction of the
small nebulae in the area  d\( _{\odot } \)~$ =  3.6$ kpc.

Clusters or stellar groups embedded in the nebulae vdBH-RN\,38 and
vdBH-RN\,53a have been identified as objects 53 and 72, respectively,
by  Dutra et al. (2003), while the clusters related to GGD\,20 and
ESO\,95-RN\,18  are given for the first time in this work.   The
cluster related to NGC\,6595 was reported in Bica et al. (2003a),
while that related to vdBH-RN\,26 was given in Dutra \& Bica (2001).

In the present study, we analyse the stellar content   of reflection
nebulae by means of the radial distribution of surface density of
sources (main sequence  stars -- MS, pre-main sequence stars -- PMS),
colour-colour and colour magnitude diagrams  (CMDs). We investigate
the nature of these objects by analysing their  physical parameters.

The cluster equatorial and galactic coordinates are shown in Table
1. The  centers are optimized peak values from the structural analysis
(Sect. 3).

In Sect. 2 we describe the spectroscopic observations and photometric
data.  In Sect. 3 we present the methods and in Sect. 4    we provide
the  photometric results for the  most populous stellar
systems. Finally, Sect. 5 provides the concluding remarks  of this
work.

\section {Spectroscopic and photometric data}

The spectroscopic observations were carried out at Complejo
Astron\'omico El  Leoncito (CASLEO, Argentina) during three nights in
May 2002, with the 2.15m  telescope. We employed a REOSC spectrograph
containing a Tektronics CCD of  1024x1024 pixels, with pixel size
24x24 $\mu$m. The total field along the  slit was $4.7\arcmin$, with
the slit oriented in the east-west direction. We  obtained spectra
ranging from 3500 \AA~to 7010 \AA, using a grating of 300
lines/mm. The average dispersion was 140 \AA/mm. In general, 3
exposures of  20 minutes for each object were taken. Standard stars --
namely, LTT\,7379,  LTT\,3864 and CD\,32 (Stone \& Baldwin 1983) --
were used for calibration of  the observed spectra. For instrumental
calibration purposes, frames of  Cu-Ar-Ne comparison lamps were taken
between and after the object  observations, as well as bias, dome,
twilight sky and tungsten lamp  flat-fields. The spectra were reduced
at the UFRGS Astronomy Department  (Brazil) using the IRAF
package. The description of reduction procedures is  given in Piatti
et al. (2000). Fig. 1 shows the observed spectrum of the  brightest
star in three of the clusters. The stars   NGC\,6595\_1,
vdBH-RN\,38\_1 and  ESO\,95-RN\,18\_1 are reddened early-type stars
appearing to be related to  the nebulae. The stars GGD20\_1   and
vdBH-RN\,26\_1, located on the periphery  of the objects,  appear to
be  unreddened late-type field stars.

\begin{figure}

\resizebox{\hsize}{!}{\rotatebox{270}{\includegraphics{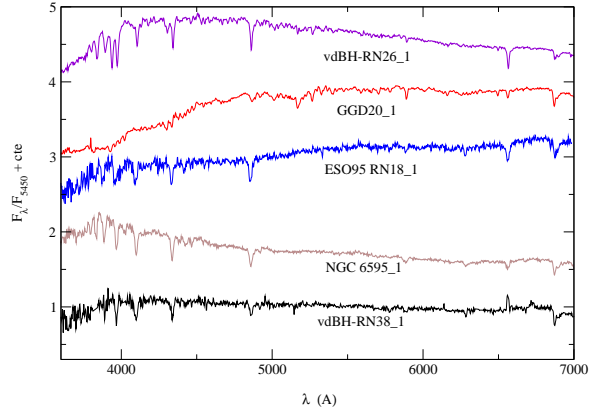}}}

\caption[]{Observed spectrum of the brightest star in five clusters.
Spectra are normalized to $F_{\lambda} = 1$ at $\lambda = 5450$ \AA.
Constants have been added to the spectra, except for the bottom one.}

\label{fig1}

\end{figure}

As photometry, we employed the Gator tool for Point Source Catalogue
extractions, as given in the Two Micron All Sky  Survey (2MASS) site
({\it http://www.ipac.caltech.edu/2mass/}). An extraction  table gives
for each star the J, H and $K_s$ magnitudes, the three  corresponding
colours ($J-H$, $H-K_s$ and $J-K_s$), the respective errors  and
J2000.0 coordinates. The 2MASS photometric errors in the present
regions  for each magnitude are basically the same as those presented
in Soares \&  Bica (2002) for the regions of NGC\,2327 and
BRC\,27. The errors become  important (0.1 mag and larger) for
magnitudes fainter than $K_s = 14$. For  comparison purposes we made
extractions up to the radius r = 5\arcmin~around  the object center.

\begin{table}

\caption[]{Assumed cluster coordinates.}

\label{tab1}

\begin{tabular}{lccccc}

\hline\hline

Cluster&$\alpha$(J2000) & $\delta$(J2000) & l($\degr$) &b($\degr$) \\
\hline

vdBH-RN\,26&08\( ^{h} \)58\( ^{m} \)04\( ^{s} \)& -47\( ^{\circ }
\)22\arcmin 54\arcsec&267.72&-1.10\\

vdBH-RN\,38&10\( ^{h} \)32\( ^{m} \)41\( ^{s} \)& -61\( ^{\circ }
\)37\arcmin 28\arcsec&287.21&-3.06\\

vdBH-RN\,53a& 11\( ^{h} \)45\( ^{m} \)57\( ^{s} \)& -65\( ^{\circ }
\)33\arcmin 34\arcsec&296.22&-3.54 \\

GGD\,20~~&7\( ^{h} \)24\( ^{m} \)41\( ^{s} \)& -24\( ^{\circ }
\)34\arcmin  42\arcsec&238.47&4.16\\

ESO\,95-RN\,18& 12\( ^{h} \)51\( ^{m} \)22\( ^{s} \)& -63\( ^{\circ }
\)18\arcmin 18\arcsec&302.92&-0.43\\

NGC\,6595&18\( ^{h} \)17\( ^{m} \)06\( ^{s} \)& -19\( ^{\circ }
\)51\arcmin 52\arcsec&11.42&-1.71\\

\hline\hline

\end{tabular}

\end{table}

We show optical and IR images for two of the analysed objects
(vdBH-RN\,26  and GGD\,20) to  illustrate a cluster or stellar group
associated with a reflection nebula.  The infrared band Digitized Sky
Survey images  (\it{http://cadcwww.dao.nrc.ca/cadcbin/getdss}\rm) of
vdBH-RN\,26 and GGD\,20 are given in  Fig. 2. The reflection nebulae
are shown in the central parts of these optical  images. Fig. 3 shows
the $ K_s$ band 2MASS images of the same objects.

\section  {Methods of analysis}

\subsection {Spectroscopy}

Spectroscopic data were obtained for the brightest star in each
object, except for vdBH-RN\,53a. In Fig.  4 we show the observed
spectrum of the vdBH-RN\,38   for the sake of illustrating the
analysis method.  We obtained spectral types for the brightest star in
each object by means of equivalent widths of Balmer lines and
comparisons with those of stars in the library of Silva \& Cornell
(1992).  This method is the same as that  employed by Ahumada et
al. (2001).  From these comparisons we also derived reddening values
(Fig. 4).  Table 2 shows the coordinates of the observed stars, the
resulting extinction values $A_V$ and the spectral types.  The
resulting extinction of the member stars are considerable for the
optical range, but not for the near infrared. The brightest source in
NGC\,6595, vdBH-RN\,38 and ESO\,95-RN\,18 is a B3-4 star, indicating
the absence of high mass stars in the cluster. This does not  produce
significant ionization, in agreement with the occurrence of a
reflection nebula. The bright star GGD\,20\_1 and vdBH-RN\,26\_1
proved to be  foreground K4 and F6-7 stars, respectively.  In the case
of vdBH-RN\,38\_1 a weak $H_\alpha$ emission line is present (Fig. 1),
but it is not clear whether it is circumstellar or nebular. In the CMD
analyses, we use the information on spectral types as constraints for
distances and, when available, other fit parameters. Otherwise, we
will use luminosity and mass constraints, aware of the fact that we
are dealing with reflection nebulae.

\begin{figure*}

\resizebox{\hsize}{!}{{\includegraphics{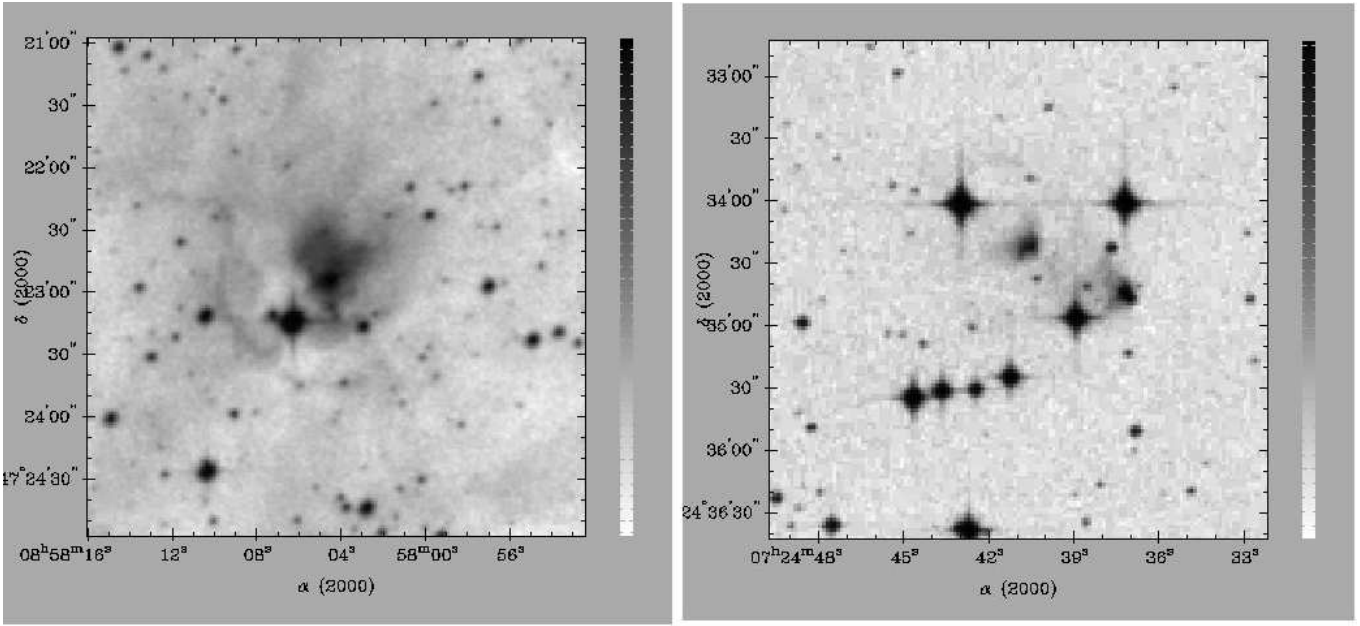}}}

\caption[]{R image of vdBH-RN\,26 and GGD\,20  from the Digitized Sky
Survey (DSS).}

\label{fig2}

\end{figure*}

\begin{figure*}

\resizebox{\hsize}{!}{{\includegraphics{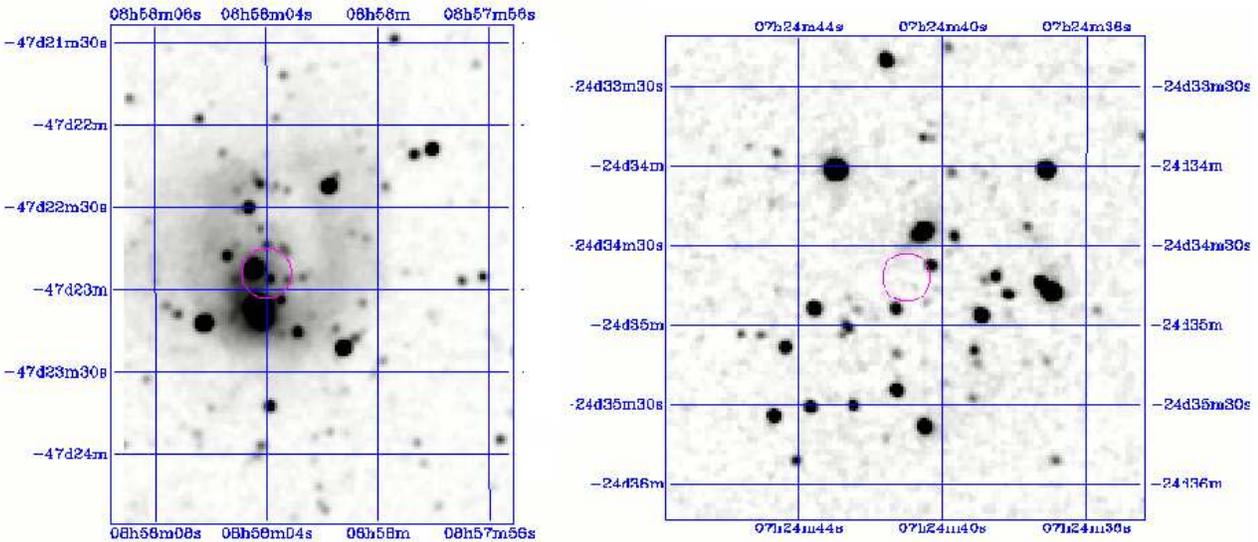}}}

\caption[]{ $K_s$ band 2MASS image of vdBH-RN\,26 and GGD\,20  showing
clustering. The small open circle indicates the object's central part.}

\label{fig3}

\end{figure*}

\begin{figure}

\resizebox{\hsize}{!}{\rotatebox{270}{\includegraphics{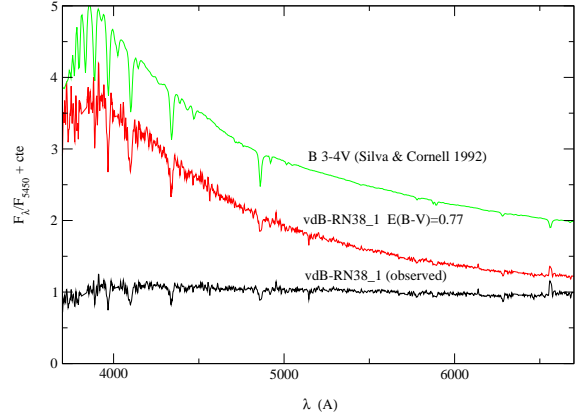}}}

\caption[]{Observed spectrum of the vdBH-RN\,38 brightest star, the
same  spectrum corrected of reddening and the B3-4 template spectrum
(Silva \&  Cornell 1992). The spectra are normalized at $\lambda =
5450$ \AA~and  shifted by arbitrary constants for comparison purposes.}

\label{fig4}

\end{figure}

\begin{table}

\caption[]{ Positions and spectroscopic results for the brightest
stars in the nebula direction.}
\label{tab1}

\begin{tabular}{lccccc}

\hline\hline

Brightest star& $\alpha$(J2000) & $\delta$(J2000) & $A_V$ & ST\\ \hline

vdBH-RN\,26\_1& 08\( ^{h} \)58\( ^{m} \)06\( ^{s} \) & -47\( ^{\circ }
\)23\arcmin 12.1\arcsec&0.2&F6-7\\

vdBH-RN\,38\_1& 10\( ^{h} \)32\( ^{m} \)40.8\( ^{s} \) & -61\( ^{\circ
} \)37\arcmin 27.1\arcsec&2.4&B3-4\\

GGD\,20\_1~~&7\( ^{h} \)24\( ^{m} \)42.9\( ^{s} \) & -24\( ^{\circ }
\)34\arcmin 01.2\arcsec &0.3&K4\\

ESO\,95-RN\,18\_1& 12\( ^{h} \)51\( ^{m} \)21.7\( ^{s} \) & -63\(
^{\circ } \)18\arcmin 11.5\arcsec&3.4&B3-4\\

NGC\,6595\_1& 18\( ^{h} \)17\( ^{m} \)05.7\( ^{s} \) & -19\( ^{\circ }
\)51\arcmin 52.5\arcsec&1.8 &B3-4\\

\hline\hline

\end{tabular}
\begin{list}{}
\item {}
\end{list}
\end{table}

\subsection{Structure}

We extracted stars within  5\arcmin~of the assumed center of each
object  (Table 1), using 2MASS. By dividing this area into concentric
rings  0.2\arcmin~wide, we calculated the number of stars confined in
each ring  per unit area. Fig. 5 displays the relation between source
surface density  and angular radial distance for the four
clusters. The indicated bars are  Poissonian errors. The dashed line
indicates the mean star field  density  which is obtained by the stars
contained in the 10 outermost rings.

\begin{figure*}

\resizebox{\hsize}{!}{\rotatebox{0}{\includegraphics{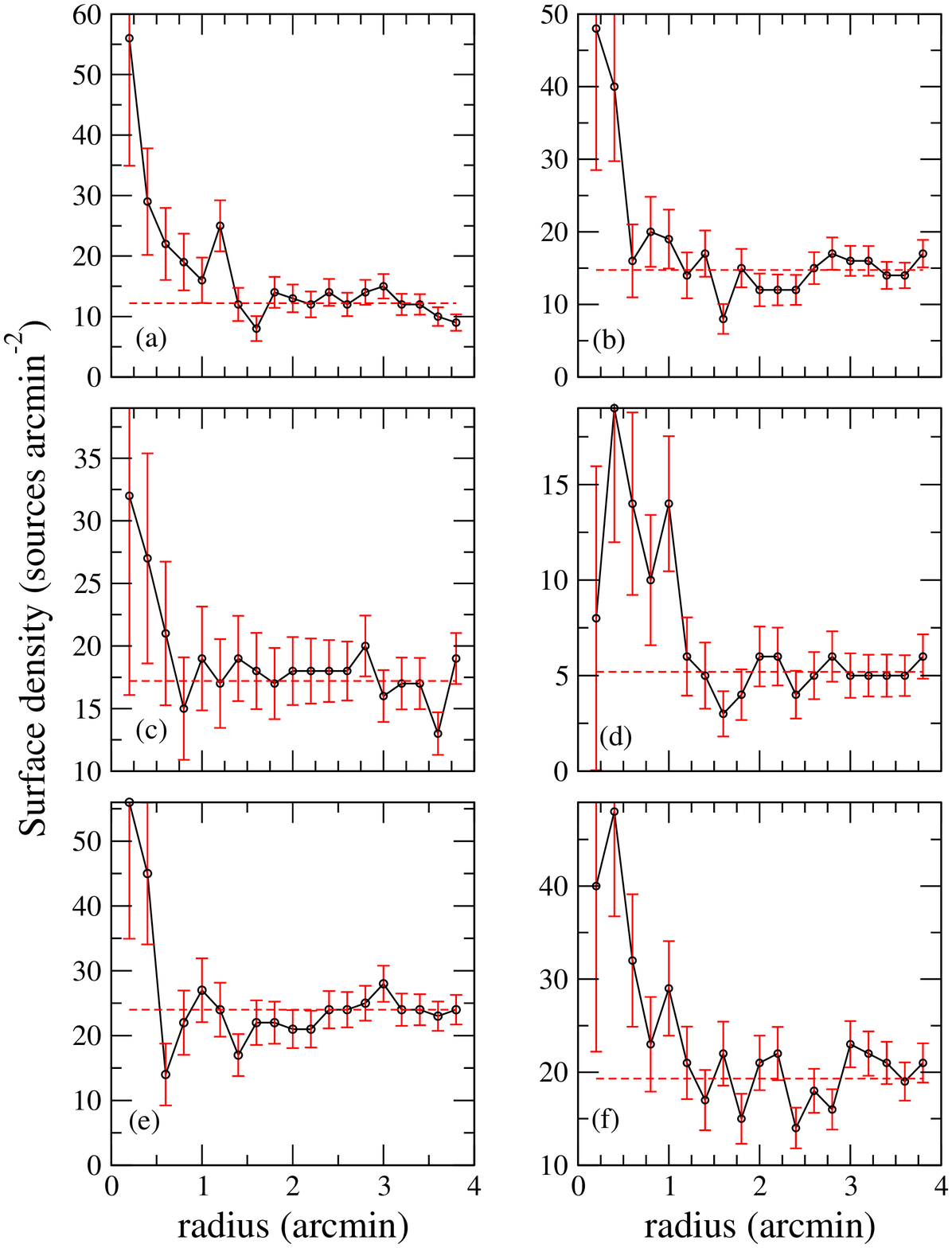}}}

\caption[]{Source surface density distribution for the studied objects
in  the nebulae:   (a)  vdBH-RN\,26, (b) vdBH-RN\,38, (c)
vdBH-RN\,53a, (d) GGD\,20,  (e)  ESO\,95-RN\,18 and  (f)
NGC\,6595. Dashed line indicates the mean background surface  density
for each object. Poissonian error bars are indicated.}

\label{fig5}

\end{figure*}

In Fig. 5 we note important source surface density contrasts for the
objects  as compared to the  background. The densities decrease below
the mean field  values for radii just outside the assumed object
radius. This can be an  indication of an underestimation of the number
of stars contained in the  cluster due to dust absorption from the
reflection nebula. We verified that  the profiles cannot be fitted by
King's law, certainly as a result of the  early evolutionary stage of
the systems.

A richness indicator index which subtracts the foreground/background
contamination was utilized  (Testi et al. 1998). In Table 3, we show
the  index for the present clusters.   Those clusters in the nebulae
vdBH-RN\,26, vdBH-RN\,38,  GGD\,20 and NGC\,6595 have clustering
values comparable to  those of the richer objects analysed by Testi et
al. (1998).

\begin{table*}

\caption[]{Structural and physical parameters for the objects.}

\label{tab1}

\begin{tabular}{lccccccccc}

\hline\hline

Cluster &  Ic$^{\mathrm{a}}$ & $A_J$$^{\mathrm{b}}$    & d
(kpc)$^{\mathrm{c}}$ & r($\arcmin$)$^{\mathrm{d}}$ &
r(pc)$^{\mathrm{e}}$ & t$\arcmin$(Myr)$^{\mathrm{f}}$ &
t$\arcsec$(Myr)$^ {\mathrm{g}}$ & $\sigma$(Myr)$^{\mathrm{h}}$ \\
\hline

vdBH-RN\,26& 44& $1.5$ & 1.7 & 1.4& 0.7 &  1-2 & $1.0$  & $1.9$  \\

vdBH-RN\,38& 22& $1.0$ & 1.2 & 1.0& 0.35 &  $>3$ &$ 2.2$ &$ 3.3$ \\

vdBH-RN\,53a& 6& - &1.7 & 0.7& 0.35 & $>3$ & -  & - \\

GGD\,20~~&24& $1.1$  & 1.3& 1.0&0.45 & 2-3 & $1.7$  & $3.1$\\

ESO\,95-RN\,18& 10& -  & 1.6& 0.5 &0.2 & $>3$ & -  & - \\

NGC\,6595 & 22 & $1.7$ & 0.6 & 1.0& 0.2 &  $>3$ & $3.7$ & $3.9$ \\

\hline\hline

\end{tabular}

\begin{list}{}
\item {$^{\mathrm{a}}$clustering index (Sect. 3); $^{\mathrm{b}}$$A_J$
mean absorption; $^{\mathrm{c}}$distance from the Sun;
$^{\mathrm{d}}$angular  radius;  $^{\mathrm{e}}$linear radius;

 $^{\mathrm{f}}$age obtained by means of the IR excess method;
$^{\mathrm{g,h}}$mean age and dispersion obtained by means of the PMS
isochrone  method.}
\end{list}

\end{table*}

\subsection{Colour-colour diagram}

Stars at formation stages (T\,Tauri and Herbig AeBe) can exhibit
anomalous  colours due to the $K_s$ band excesses created by  hot dust
emission  around them (Carpenter et al. 1993; Lada \& Adams 1992;
Calvet et al.  1992), an effect which is observed in the $((J-H),
(H-K_s))$ diagram. By  means of the percentage of anomalous stars, one
can estimate the cluster or  stellar group age (Lada et al. 1996,
Soares \& Bica 2003). This diagram also  provides the mean reddening
of the object.

Fig. 7 displays a $((J-H), (H-K_s))$ diagram, which shows intrinsic
colours  of MS stars from O3 to M5 (Binney \& Merrifield  1998 -- and
references  therein), together with reddening lines of
$E(J-H)/E(H-K_s)=1.72$  (Schlegel et al. 1998) for an O3 MS star and
an M5 giant, as well as the  locus of unreddened T\,Tauri stars (Meyer
et al. 1997).

Extinction is estimated from the observed colours in the $((J-H),
(H-K_s))$  diagram. The intrinsic colours of the embedded stars should
be known. We  consider the following cases: (i) the  domain of normal
reddening containing  weak T\,Tauri stars, background/foreground stars
and highly reddened  luminous early-type PMS stars (Strom et
al. 1995); (ii) the domain of  classical $K_s$ and $L$ band excesses,
from the reddening line of O3  stars to the reddening line of the
reddest colour reached by standard disk  models. This domain contains
classical T\,Tauri stars; (iii) the domain of  objects with larger
$K_s$ and $L$ band excesses located to the right of the  reddest
colour predicted for a standard disk model (Lada \& Adams 1992).
Stars contained in this domain are probably surrounded by extended
envelopes  (Kenyon et al. 1993) or heavily reddened Herbig AeBe stars
with central  holes in their disks (Lada \& Adams 1992; Hillenbrand et
al. 1992).

The stars contained in domain (iii) were not dereddened because their
intrinsic colours are not straightforwardly obtained, as a consequence
of a  wide range of colours produced by their surrounding envelopes
(Strom et al.  1995). Extinction was obtained only for stars located
in domains (i) and  (ii), using reddening vectors and intrinsic
colours.

  \subsection {Subtraction method of field stars}

The field stars were removed from the colour-colour and
colour-magnitude diagrams   based on an analysis of the colour-colour
diagram.   Comparisons of the distribution of stars on the latter
diagram for on-cluster and off-cluster stars are  the essence of the
statistical field subtraction method applied.  The off-cluster and
on-cluster regions are concentric. The off-cluster region corresponds
to $ 2\arcmin < r < 6\arcmin$. The method is similar  to that
presented in Kerber  et al. (2002) for CMDs of a Magellanic Cloud
cluster. It is based on the hypothesis that  positions of  off-cluster
stars represent the most likely positions for field stars on similar
colour-colour diagram of the surroundings. We estimate the probability
of each on-cluster star being a cluster member, by  the numbers of
stars of the  on-cluster and the off-cluster contained in a
$3\sigma_{H-K_s}\times3\sigma_{J-H}$ box centered in it.

The probability $P_J$ that the $j$th on-cluster star is any  of the
$N_{on} $ member stars is:

\vspace{.5cm}

$P_{j}  = 1 - n_{off} / n_{on} \times \Omega_{on}/\Omega_{off}$,

\vspace{.5cm}

where $\Omega_{on} $ and $ \Omega_{off}$ are the solid angles of the
respective regions.

The sum of $P_{j}$ over the  $j$th on-cluster star is  $N_{on}$, the
number of cluster member stars pointed out by the method.

According to this probability we randomly remove stars from the
 on-cluster colour-colour diagram. In Fig. 6  we have a comparison of
 an observed and a field subtracted  colour-colour diagram.

\begin{figure}

\resizebox{\hsize}{!}{\rotatebox{0}{\includegraphics{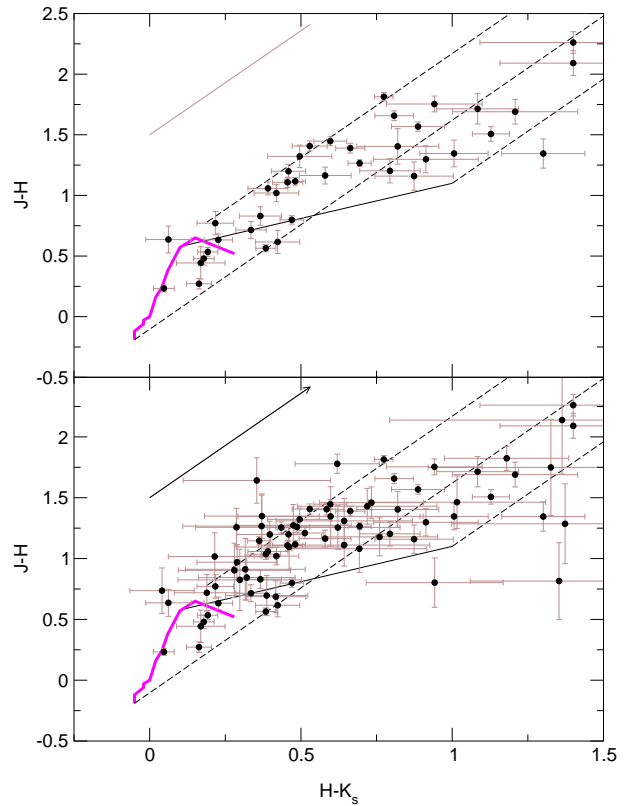}}}

\caption[]{Example of a statistical  field star subtraction on the
colour-colour diagram:  upper panel  is the subtraction result, while
the lower panel is the observed one. The object is vdBH-RN\,26.  }

\label{fig6}

\end{figure}

\subsection {Colour magnitude diagram}

We analysed the $(J,(J-H))_0$ CMDs by means of PMS evolutionary tracks
and isochrones of D'Antona \& Mazzitelli (1997, 1998). The
evolutionary tracks  are from 0.07 Myr to 100 Myr for PMS stars in the
range of 0.17 M\( _{\odot }  \) to 3 M\( _{\odot } \). The deuterium
abundance adopted is  $2\times10^{-5}$ and metallicity is $Z=0.02$. We
converted the theoretical  into observational plan by using the
bolometric corrections of  Schmidt-Kaler (1982). The 0.3, 1, 3 and 10
Myr isochrones, together with a  zero age main sequence (Bertelli et
al. 1994) and evolutionary tracks for  3, 2, 1 and 0.5 M\( _{\odot }
\), are shown in Fig. 8.

  In the CMD analysis it is fundamental to decontaminate it for field
 stars. The member stars obtained in Sect. 3.4 for each cluster were
 used in the present analysis.  The clusters in the nebulae
 vdBH-RN\,26, vdBH-RN\,38,  GGD\,20 and NGC\,6595 have good statistics
 and their  analysis  is given in Sect. 4. vdBH-RN\,53a and
 ESO\,95-RN\,18  showed too few stars and will not be further
 discussed. However the density profiles (Sect.3.2) suggest  that they
 are stellar systems. Thus, we strongly recommend observation with
 larger telescopes.

 We estimated the  observed mass of the clusters from the dereddened
CMDs, using the  evolutionary tracks. The observed cluster masses
proved to be smaller than  $M_t=25M_{\odot}$ for each cluster. This
value can be considered  an underestimation of the total mass, owing
to the photometric cutoff and  to the presence of dust  in the objects.

{\bf As a second method, the object ages were estimated by using the
PMS  isochrones.  An  age fitting was applied to  each  member star in
the CMD, previously corrected for absorption (Sect. 3.3), then we
derived ages and age dispersions  for the clusters.
 
The age and dispersion  errors were obtained  from the values  found
for twenty realizations of the statistical subtraction.

Table 3 gives the age  t$\arcsec$ obtained by means of the PMS
isochrone method for each cluster  and the respective dispersion
$\sigma$. For PMS stars above the 0.3 Myr isochrone,  we adopted this
lower limit age.  Some field contamination can be present in the
cluster CMDs, despite  their low membership probability.  Some of
these stars with resulting ages older than 10 Myr were not included in
the mean value calculations. The age dispersions  are non-negligible
and compatible with expected dispersions for the cluster  formation
stage. For comparison purposes, the ages t$\arcmin$ obtained by  means
of the $K_s$ band excess method (Sect. 3.3) are also shown. In
general,  a good agreement occurs.}

\section {Discussion of the individual clusters}

\subsection {The cluster in the nebula vdBH-RN\,26}

The source surface density distribution for the vdBH-RN\,26 cluster
shows  an excess up to r = 1.4\arcmin \, (Fig. 5). We adopted this
radius for the  object.

\begin{figure*}

\resizebox{\hsize}{!}{\rotatebox{270}{\includegraphics{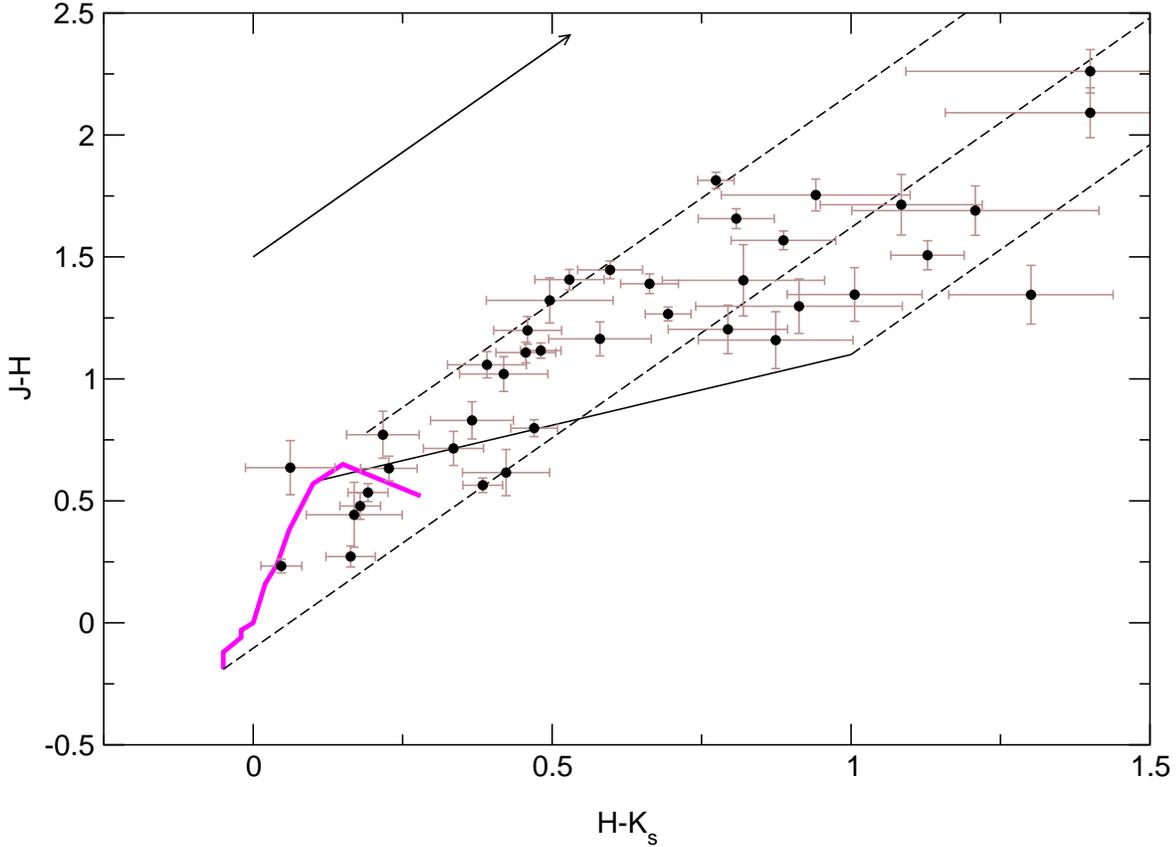}}}

\caption[]{ vdBH-RN\,26: ($(J-H),(H-K_s)$) diagram for the cluster (filled 
circles). The continuous curve represents the intrinsic distribution
of  spectral types and the continuous straight line is the unreddened
locus of T\,Tauri stars. The reddening vector represents
$A_V=5$. Reddening  lines for M5 giants, O3 stars and T\,Tauri stars
are shown as dashed lines.  Faint sources with prohibitive errors were
excluded. Photometric errors are indicated. The diagram corresponding to
a given  statistical  subtraction, according to the probabilities in 
Sect. 3.4.}

\label{fig7}

\end{figure*}

Fig. 7 shows, as an  example, one of  the colour-colour diagrams
 obtained from the  several realizations of the field subtraction
 statistical method.  The diagram  presents a large number of stars
 outside the first domain (Sect 3.2).  Considering the number of stars
 predicted for  the cluster, nearly 40\% of them have IR excess, such
 as in the case of the YSOs in the Taurus dark clouds, which have  an
 estimated  age of about  1-2 Myr  (Kenyon \& Hartmann 1995).  The
 mean age obtained by means of the PMS  isochrone is $1,0\pm0.2$ Myr,
 which is adopted and the age dispersion  is $1.9\pm0.5$ Myr.  The mean
 reddening  is $A_J= 1.5\pm0.1$ considering the stars in the  domain
 where intrinsic colours are known. This reddening value corresponds
 to $A_V = 5.6\pm0.3$. The relations $A_J=2.6E(J-H)$ and
 $A_J=0.276A_V$, used in reddening transformations, were derived from
 the data in Schlegel et al.  (1998), assuming $A_V = 3.1E(B-V)$
 (Cardelli et al. 1989).  A CMD obtained for  vdBH-RN\,26 as a
 realization  of a statistical subtraction   is shown in Fig. 8. The
 best fit of PMS tracks to the corresponding observed objects is given.

We obtained a distance of d\( _{\odot } \)~$ = 1.7$ kpc (Sect. 1) for
the object, in agreement with van den Bergh \& Herbst (1975).

The second brightest star $((J-H)_0 \approx0.2)$ is an F6-7 foreground
  star (Table 2). The star with a similar colour located below the
  ZAMS is  probably also a field star.  One has to keep in mind that
  the method does not eliminate  all  field stars in a given
  realization. However, the effects of such low probability stars will
  be small on the mean values of the studied parameters.
    
 It is worth noting that the low mass stars have been formed in the
 cluster without the presence of a massive star.

\begin{figure*}

\resizebox{\hsize}{!}{\rotatebox{270}{\includegraphics{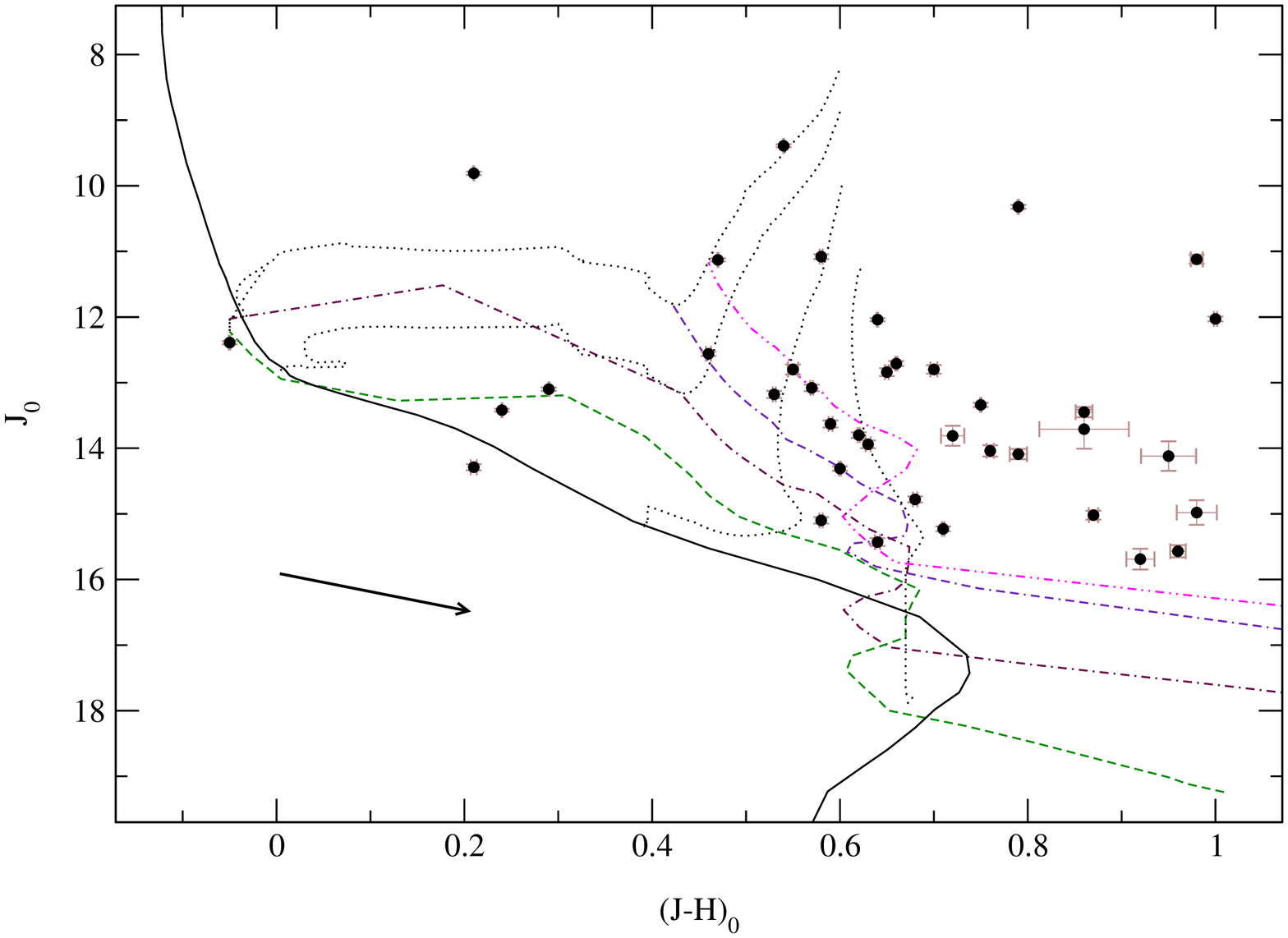}}}

\caption[]{ vdBH-RN\,26:  dereddened  $($J,(J-H)$)_0$ diagram for the cluster. The 
isochrones of 0.3 (dash), 1 (dot-dash), 3 (dot-dash-dash) and 10 Myr
(dot-dot-dash) downward are presented as dashed lines. Dotted curves
(downward) represent evolutionary tracks for 3, 2, 1 and 0.5 M\(
_{\odot }  \) (D'Antona \& Mazzitelli 1997, 1998). The ZAMS is
represented by the  continuous line (Bertelli et al. 1994). The
reddening vector corresponds to  $A_V=2$.  The  star colours are dereddened
individually. Colour and magnitude error bars  are indicated. Same 
statistically subtracted sample as in Fig. 7. }

\label{fig8}

\end{figure*}

\rm

\subsection {The cluster in the nebula vdBH-RN\,38}

  \rm

There occurs an excess of star surface density up to $\approx1\arcmin$
as  compared to the surrounding field (Fig. 5). We adopt a limiting
radius r =  1\arcmin~ for the subsequent analysis.

\begin{figure}

\resizebox{\hsize}{!}{\rotatebox{270}{\includegraphics{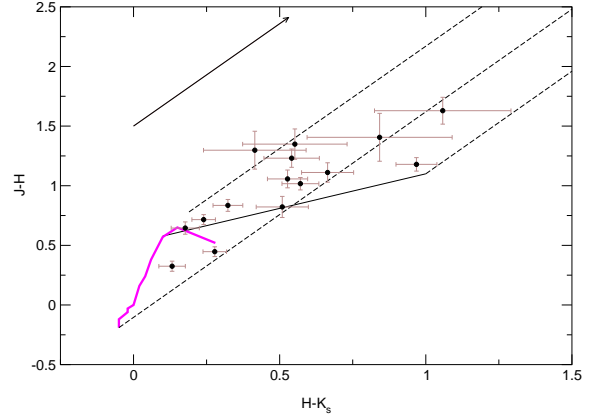}}}

\caption[]{vdBH-RN\,38: ($(J-H),(H-K_s)$) diagram for the cluster. 
Symbols and statistical subtraction as in Fig. 7. }

\label{fig9}

\end{figure}

The mean reddening is $A_J = 1.0\pm0.1$,  which corresponds to $A_V=
3.8\pm0.4$. vdBH-RN\,38 is not heavily reddened as  compared to
massive embedded clusters (Lada \& Lada 2003).  The photometric  $A_V$
value is somewhat larger than that obtained spectroscopically for  the
brightest star (Table 2). This suggests that the dust column in the
direction of the brightest star is smaller than those of the PMS stars
on  average, possibly due to stellar winds. Some stars have IR
colours  denoting infrared excess. The vdBH-RN\,38 cluster has about
$20\%$ of  PMS stars with infrared excess, which suggests an
evolutionary stage similar  to that of IC\,348 cluster which has 2-3
Myr (Haisch et al. 2001). The  value obtained by means of the PMS
isochrones for the mean age was $ 2.2\pm0.6$ Myr, in agreement with the
value obtained with the previous method. The age dispersion found  was
$ 3.3\pm0.8$ Myr. The estimate of the observed mass is less than
$M=25M_{\odot}$.

The $(J,J-H)_0$ CMD (Fig. 10) includes the PMS isochrones and
evolutionary  tracks. The fit was based on the cluster colour-colour
diagram and on the  most luminous star, which is a B3-4 spectral type
star (Sect. 3.1). The observed distance modulus is $J-M_J= 10.4$,
implying a distance from the Sun of d\( _{\odot } \)~$ = 1.2$
kpc. This  value is a factor of $\approx$ 2 smaller than previous
estimates for the  nebula (Sect. 1). This solution is constrained in
terms of the intermediate  mass star and of  PMS objects.

\begin{figure*}

\resizebox{\hsize}{!}{\rotatebox{270}{\includegraphics[width=12cm]{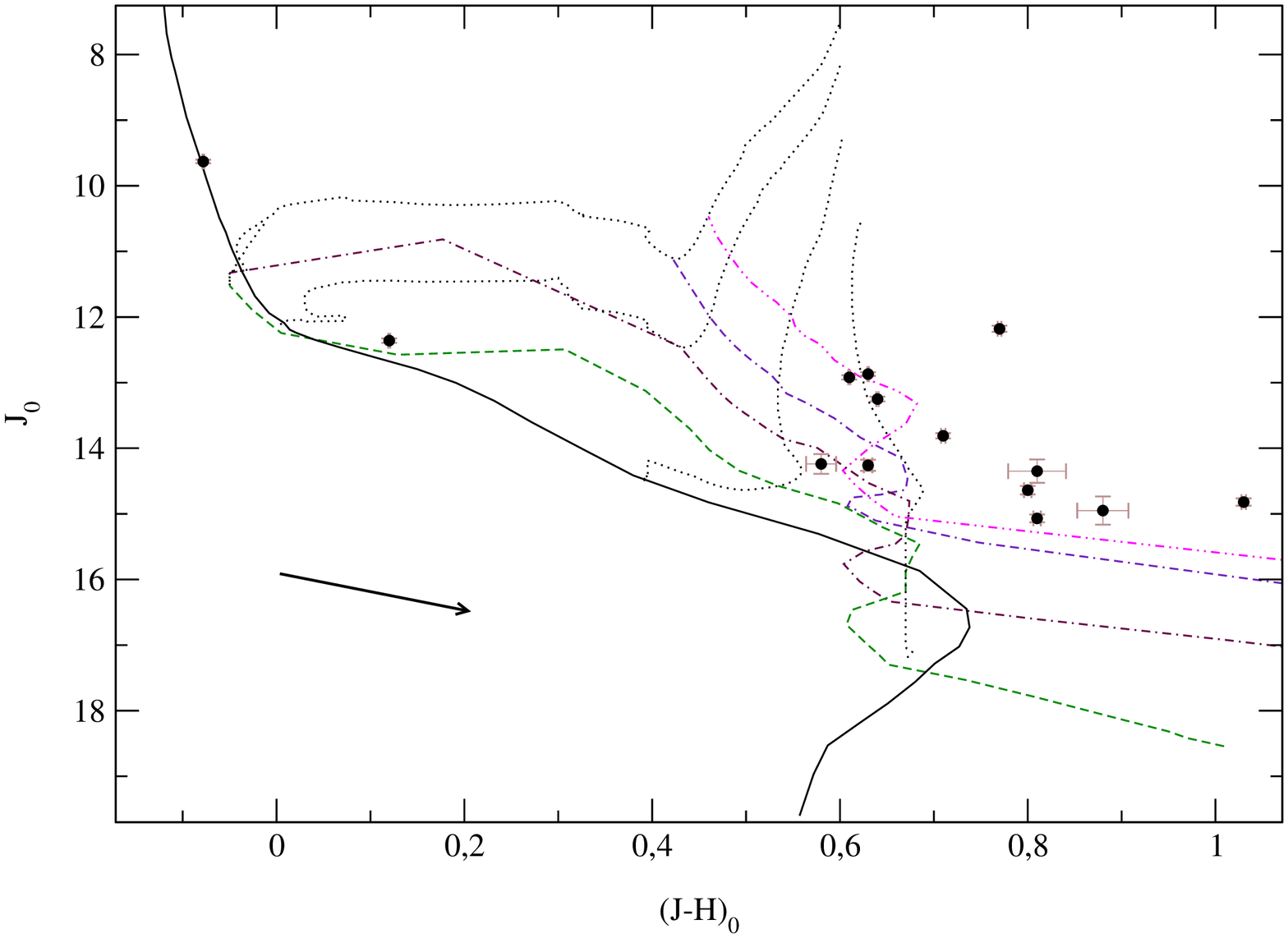}}}

\caption[]{ vdBH-RN\,38: dereddened  $(J,(J-H))_0$ diagram for the cluster.
 Symbols as in Fig. 8. Same statistically subtracted sample as in Fig. 9. }

\label{fig10}

\end{figure*}


\subsection {The cluster in the nebula GGD\,20}

In contrast to the previous objects, GGD\,20 presents a comparatively
lower  source surface density. In Fig. 5 the surface density is
considerably larger  than that of the surrounding field up to $ r =
1\arcmin$, which we adopted for extractions.

\begin{figure}

\resizebox{\hsize}{!}{\rotatebox{270}{\includegraphics{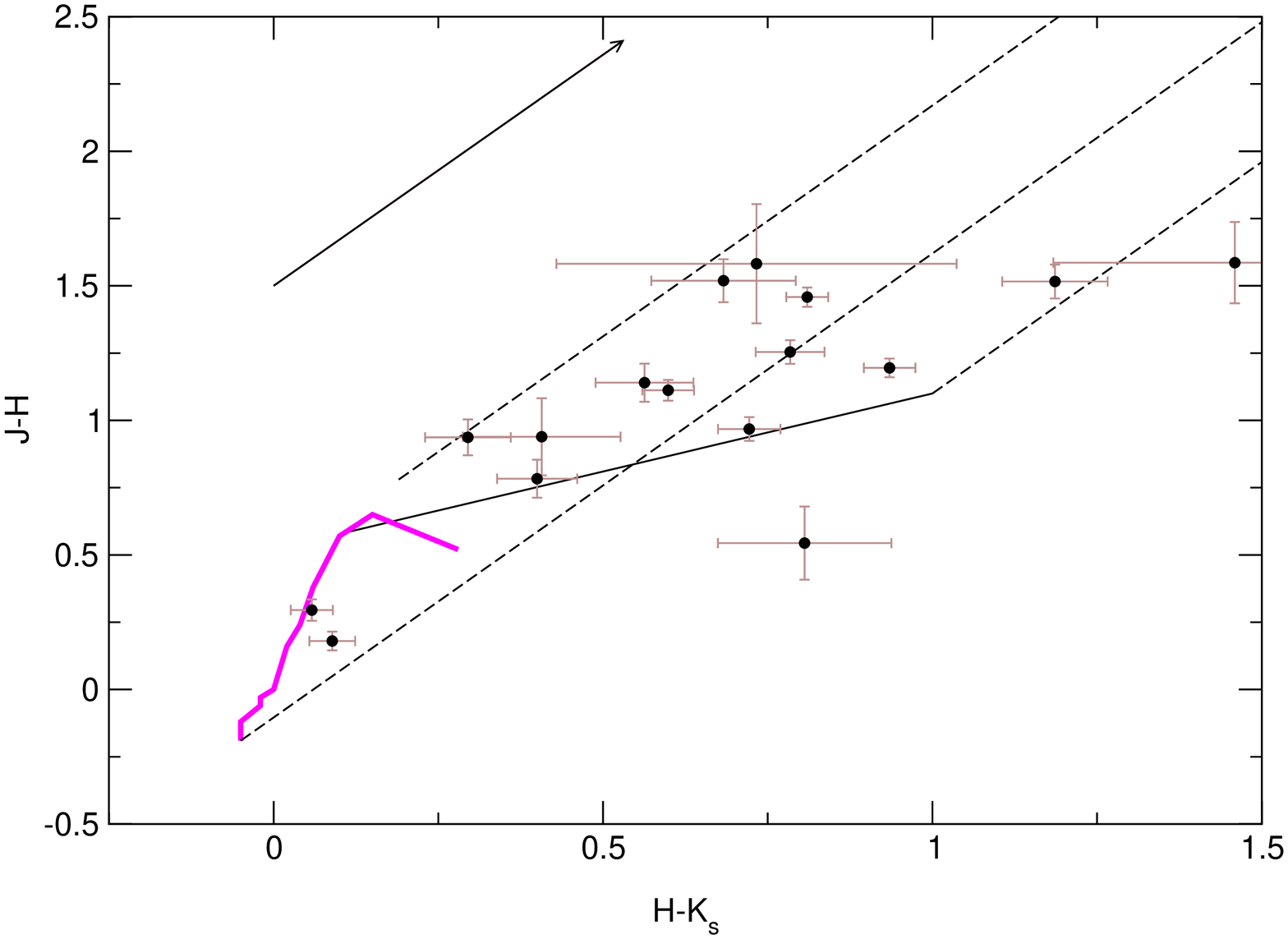}}}

\caption[]{ GGD\,20:  ($(J-H),(H-K_s)$) diagram for the cluster. Symbols and
 statistical subtraction as in Fig. 7.}

\label{fig11}

\end{figure}

The ($(J-H),(H-K_s)$ diagram (Fig. 11) shows $30\%$ of the stars with
$K_s$  band excess, implying a similar evolutionary stage to that of
the IC\,348  cluster. The PMS isochrone method gives an mean age of
$1.7\pm0.2$ Myr and a age dispersion of $3.1\pm0.3$ Myr.  The mean
reddening is $A_J= 1.1\pm0.1$ considering the stars with predictable
intrinsic colours. This reddening value corresponds to  $A_V =
4.0\pm0.5$.

In Fig. 12 we show the $(J,(J-H))_0$ diagram for the cluster, together
with  the PMS isochrones fit.  The  brightest star in the cluster
direction resulted to be a K4 star (Sect. 3) and  was discarded by the
field star subtraction. The observed distance  modulus is $J-M_J=
10.5$ and the distance is d\( _{\odot } \)~$ = 1.3$ kpc.  The triangle
in Fig. 12 represents the Herbig AeBe star described in Vieira  et
al. (2003). The most massive star near the ZAMS probably has
$M\approx$ 2M\( _{\odot }  \) and spectral type $\approx$A5. Intermediate
 brightness  stars redder than $(J-H)_0
= 0.7$ appear to be classical T\,Tauri stars  with IR emission from
surrounding disks. The star distribution along the  CMD points to a
mean age compatible to that obtained by means of the $K_s$  band
excess  method.

\begin{figure*}

\resizebox{\hsize}{!}{\rotatebox{270}{\includegraphics{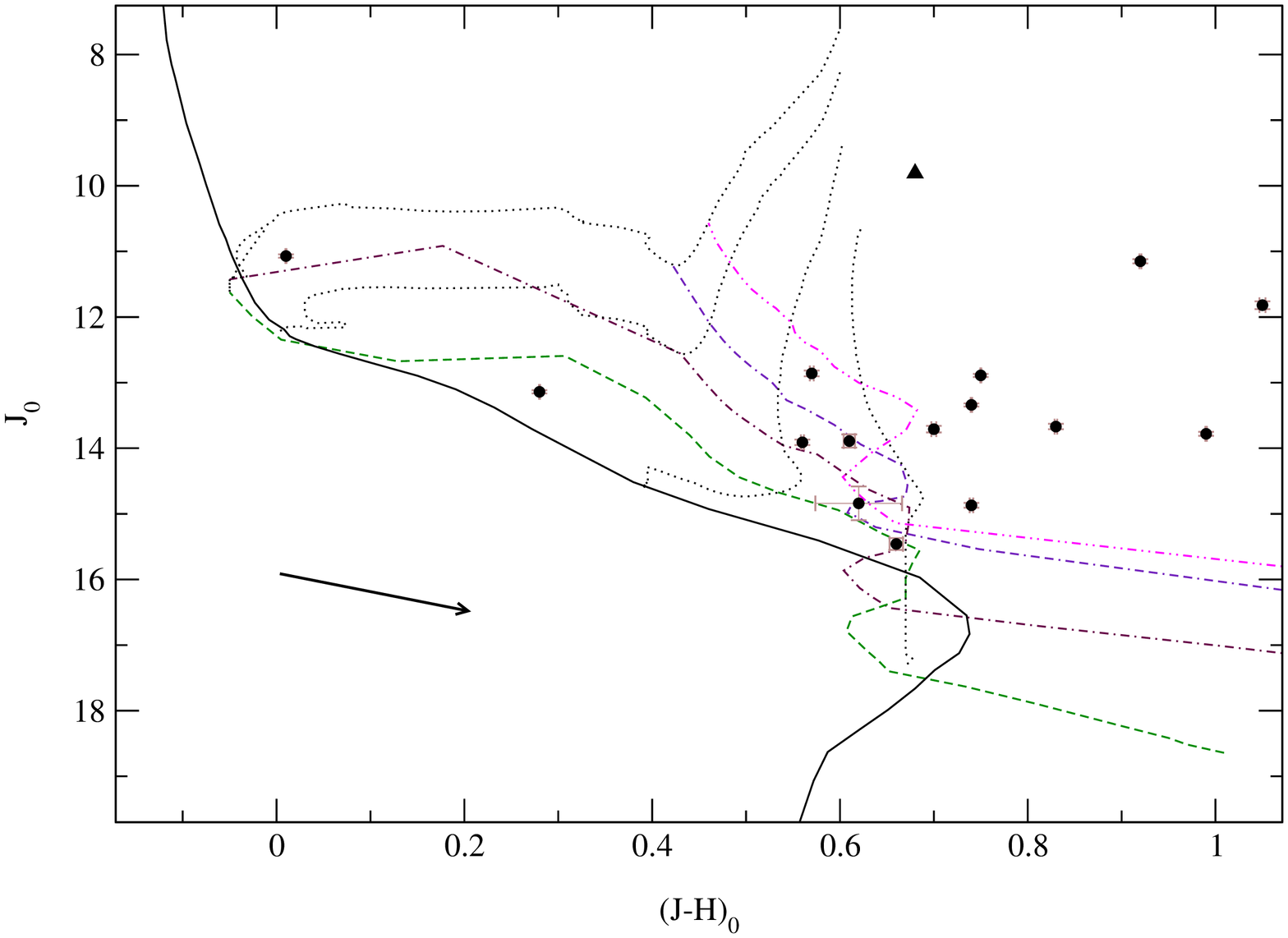}}}

\caption[]{ GGD\,20: dereddened  $(J,(J-H))_0$ diagram for the cluster. The triangle represents a Herbig AeBe star. Other symbols as  in Fig. 8. 
Same statistically subtracted sample as in Fig. 11.  }

\label{fig12}

\end{figure*}

\subsection {The cluster in the nebula NGC\,6595}

The source surface density (Fig. 5) shows a large concentration of
stars  within $ r = 1.2\arcmin$. The angular radius assumed in the
analysis for the NGC\,6595 cluster is $ r = 1\arcmin$.

\begin{figure}

\resizebox{\hsize}{!}{\rotatebox{270}{\includegraphics{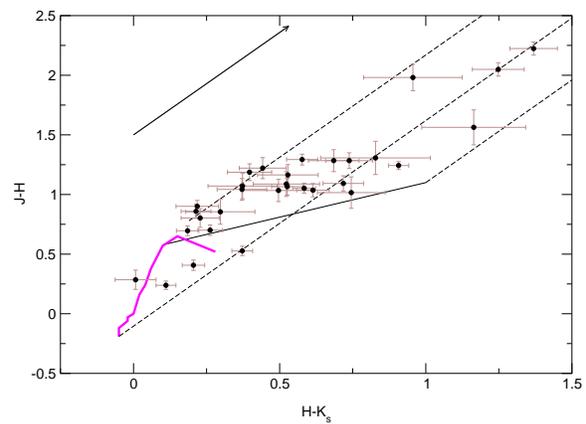}}}

\caption[]{ NGC\,6595:  ($(J-H),(H-K_s)$) diagram for the cluster. 
Symbols and statistical subtraction as in Fig. 7. }

\label{fig13}

\end{figure}

A large number of stars in the direction of the nebula  NGC\,6595  are
red giants from the bulge. These  stars were removed in the
statistical star field subtraction. A $((J-H),(H-K_s))$ diagram,
resulted from the field subtraction method, is shown in Fig. 13.

The percentage of the cluster stars with $K_s$ band excess is 20\%,
similar to that of the vdBH-RN\,38 cluster.

The mean reddening is $A_J = 1.7\pm0.1$ ($A_V= 6.1\pm0.3$). This value
 is somewhat larger  than the spectroscopic one (Table 2) -- see
 discussion in Sect. 4.1.  The PMS isochrone  method indicates an age
 of $3.7\pm0.2$ Myr and an age dispersion of  $3.9\pm0.1$ Myr.

In Fig. 14 we show an obtained $(J,(J-H))_0$ CMD for the object
together with the PMS  isochrone fit.  Spectroscopic  data for the
most massive star indicates a B3-4 spectral type (Sect. 3.1). The fit
points  to an object that presents most of the stellar masses in the
range  $\approx$0.5M\( _{\odot } \) to $\approx$1M\( _{\odot } \). The
observed  distance modulus is $J-M_J= 9.1$, implying a distance from
the Sun of d\( _{\odot } \)~$ =  0.6$ kpc.

\begin{figure*}

\resizebox{\hsize}{!}{\rotatebox{270}{\includegraphics{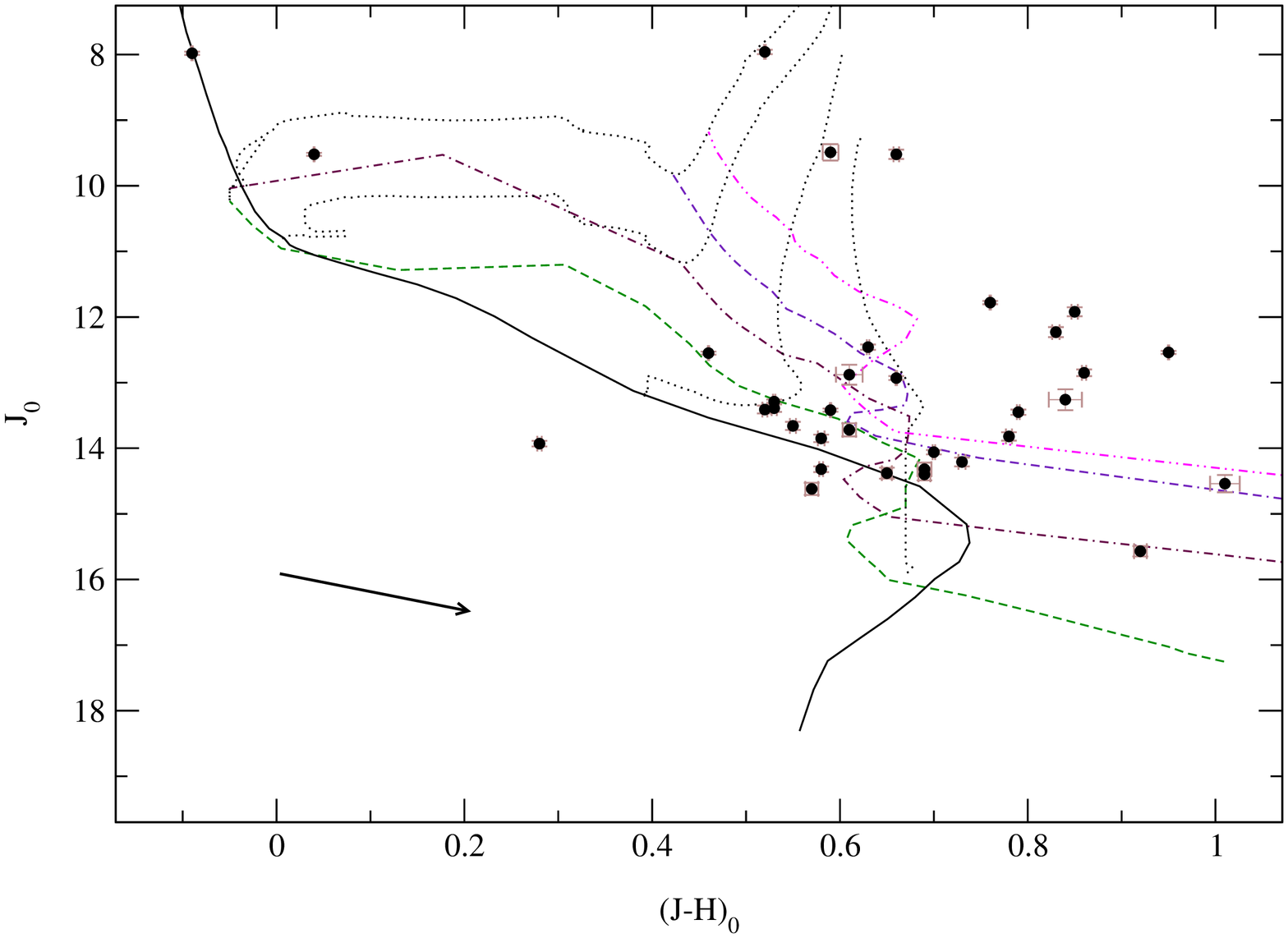}}}

\caption[]{ NGC\,6595: unreddened $(J,(J-H))_0$ diagram for the cluster.
 Symbols as in Fig. 8. Same statistically subtracted sample as in Fig. 13.}

\label{fig14}

\end{figure*}

\section{Concluding remarks}

Six candidate stellar systems were investigated by means of optical
spectroscopy and  infrared photometry in the reflection nebulae
vdBH-RN\,26, vdBH-RN\,38, vdBH-RN\,53a,  GGD\,20, ESO\,95-RN\,18 and
NGC\,6595. Except for vdBH-RN\,53a and ESO\,95-RN\,18,  which are less
populous in stars and require larger telescope data, we carried out an
analysis of the CMDs and colour-colour diagrams.

The distance found for the  vdBH-RN\,26 cluster is  d\( _{\odot } \)~$
 = 1.7$ kpc,  coincident with previous estimate in the literature for the
 nebula (Sect. 1). For vdBH-RN\,38 we found  d\( _{\odot } \)~$ = 1.2$ kpc, 
closer than previous estimates.  These objects are in different
 evolutionary stages. The larger fraction of $K_s$ excess stars in
  vdBH-RN\,26 is an indication
 of a young object, confirmed by the mean age value $1.0\pm0.2$ Myr, found
 based on PMS isochrones. vdBH-RN\,38  has  $2.2\pm0.6$ Myr. The
 reddening values are $A_V=5.6\pm0.3$  and $A_V=3.8\pm0.4$,
 respectively, considerably smaller than those found for  deeply
 embedded clusters in molecular clouds, where $A_v$ can be as large as
$\approx$100 (Lada \& Lada 2003).

The GGD\,20 cluster has $\approx$30\% of its stars with infrared
excess,  suggesting an age of 2-3 Myr, a somewhat larger than the mean
age  $1.7\pm0.2$ Myr found by means of the PMS isochrones method. The
brightest star in the nebula direction is a foreground K4 star. The
distance is  d\( _{\odot } \)~$ =  1.3$ kpc, and the mean absorption
is $A_V=4.0\pm0.5$.  The  NGC\,6595 cluster  has a mean age of
$3.7\pm0.2$ Myr.  Its distance is d\( _{\odot } \)~$ =  0.6$ kpc,
while the absorption has been  found to be $A_V=6.1\pm0.3$.

These stellar systems appear to be low mass embedded stellar clusters
containing  few stars (observed cluster masses smaller then
$25M_{\odot}$). Their $A_V$  absorptions and ages point to an
evolutionary stage similar to that of the  clusters in the molecular
cloud CMaR1 (Soares \& Bica 2002, 2003). The  analysed clusters have
no stars with spectral types earlier than B,  in agreement with the
nature of the reflection nebula, without significant  ionization as in
an HII region. In particular, the central stars of  vdBH-RN\,38 and
NGC\,6595 are B3-4 stars.


The masses estimated for the present objects indicate that they are
unbounded stellar systems (Kroupa \& Boily 2002).  The number of young
clusters or groups catalogued or studied in detail is small
(Sect. 1). Since the mass enclosed in these objects is not large, they
do not appear to be major contributors to the field population of
stars.  

\begin{acknowledgements}

This publication makes use of data from the Two Micron All Sky Survey,
which  is a joint project of the University of Massachusetts and the
Infrared  Processing and Analysis Center, funded by the National
Aeronautics and Space  Administration and the National Science
Foundation. We employed data from  the CDS database (Strasbourg). We
acknowledge the referee  and  Leandro Kerber  by the important
comments on this study.  We acknowledge support from the Brazilian
institution CNPq, in particular J.S. for a CNPq PhD fellowship. This
work  was also partially supported by the Argentinian institutions
CONICET, Secyt  (Universidad Nacional de C\'ordoba), Agencia C\'ordoba
Ciencia and Agencia Nacional de promoci\'on Cient\'{\i}fica y
Tecnol\'ogica (ANPCyT).

\end{acknowledgements}

\end {document}